\documentclass{ptapap}

\author{Sean B. Begy}[RMC]
\author{Gregg A. Wade}[RMC]
\author{Gerald Handler}[CAMK]
\author{Andrzej Pigulski}[Wroclaw]
\author{James Sikora}[RMC]
\author{Matt Shultz}[Uppsala]
\author{the BRITE team}
\affil[RMC]{Department of Physics, Royal Military College of Canada, Kingston, Ontario K7K 7B4, Canada}
\affil[CAMK]{Nicolaus Copernicus Astronomical Center, Bartycka 18, 00--716 Warszawa, Poland}
\affil[Uppsala]{Department of Physics and Astronomy, Uppsala University, Box 516, Uppsala 75120}
\affil[Wroclaw]{Instytut Astronomiczny, Uniwersytet Wroclawski, Kopernika 11, PL-51-622 Wroclaw, Poland}
\title{Evolving pulsation of the slowly rotating magnetic $\beta$ Cep star $\xi^1$~CMa}

\begin{document}

\maketitle

\begin{abstract}

We report BRITE-Constellation photometry of the $\beta$~Cep pulsator $\xi^1$~CMa. Analysis of these data reveals a single pulsation period of $0.2095781(3)$~d, along with its first and second harmonics. We find no evidence for any other frequencies, limiting the value of this star as a target for magneto-asteroseismology. We employ the 17-year database of RV measurements of $\xi^1$~CMa to evaluate evidence for the reported change in pulsation period, and interpret this period change in terms of stellar evolution. We measure a rate-of-change of the period equal to $0.009(1)$~s/yr, consistent with that reported in the literature.

\end{abstract}

\section{Introduction}

$\xi^1$~CMa is a bright ($V=4.3$), single, B0.5 subgiant located late in its main sequence evolution. It has long been known to exhibit $\beta$~Cep pulsations with a period of $P=0.20958$~d \citep[e.g.][]{1953PASP...65..193M,1992A&AS...96..207H, 2006CoAst.147..109S}. \citet{2009RMxAC..36..319H} reported the star to be magnetic, and \citet{2017MNRAS.471.2286S} carried out long-term magnetic monitoring to demonstrate that the star's rotational period is remarkably long, over 30 years. \citet{2017MNRAS.471.2286S} performed a detailed determination of the star's magnetic and physical parameters, and also examined the behaviour of the radial velocity (RV) pulsations of the star over a span of 17~yr. They demonstrated that a constant pulsation period was unable to phase those data, and consequently inferred that the period was increasing at a rate of 0.0096 s/yr.

In this paper we report two-colour BRITE photometry of $\xi^1$~CMa, examine its photometric periodogram for evidence of additional pulsation frequencies, and revisit the period evolution reported by \citet{2017MNRAS.471.2286S}.

\section{Observations}

Our analysis is based in part on BRITE-Constellation photometry of $\xi^1$~CMa obtained in 2015. Observations were obtained with BRITE-Toronto (BTr; 3 setups, obtaining 64,883 observations over 157 days), BRITE-Lem (BLb; 4 setups, 41,399 observations, 167 days) and BRITE-Heweliusz (BHr; 3 setups, 55,976, 167 days). The observations were reduced and de-trended, yielding typical $1\sigma$ uncertainties of the orbit-averaged measurements of 1.5~mmag for BTr, 3.6~mmag for BHr, and 5.9~mmag for BLb. For this preliminary analysis, we focus on the highest-quality BTr data.

\section{Period analysis}

\begin{figure}
  \centering
  \begin{minipage}{\textwidth}
    \includegraphics[width=0.45\textwidth]{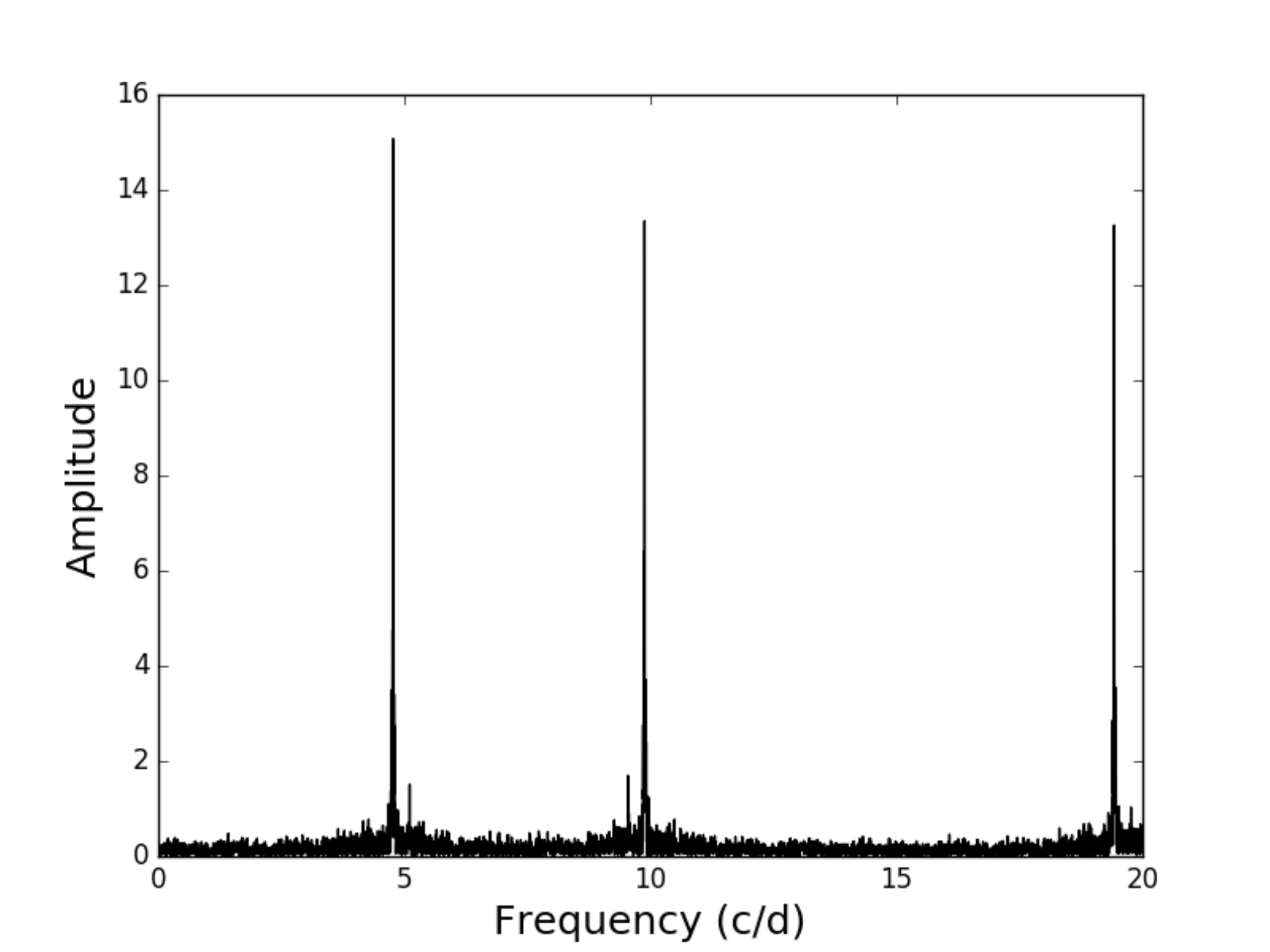}\includegraphics[width=0.45\textwidth]{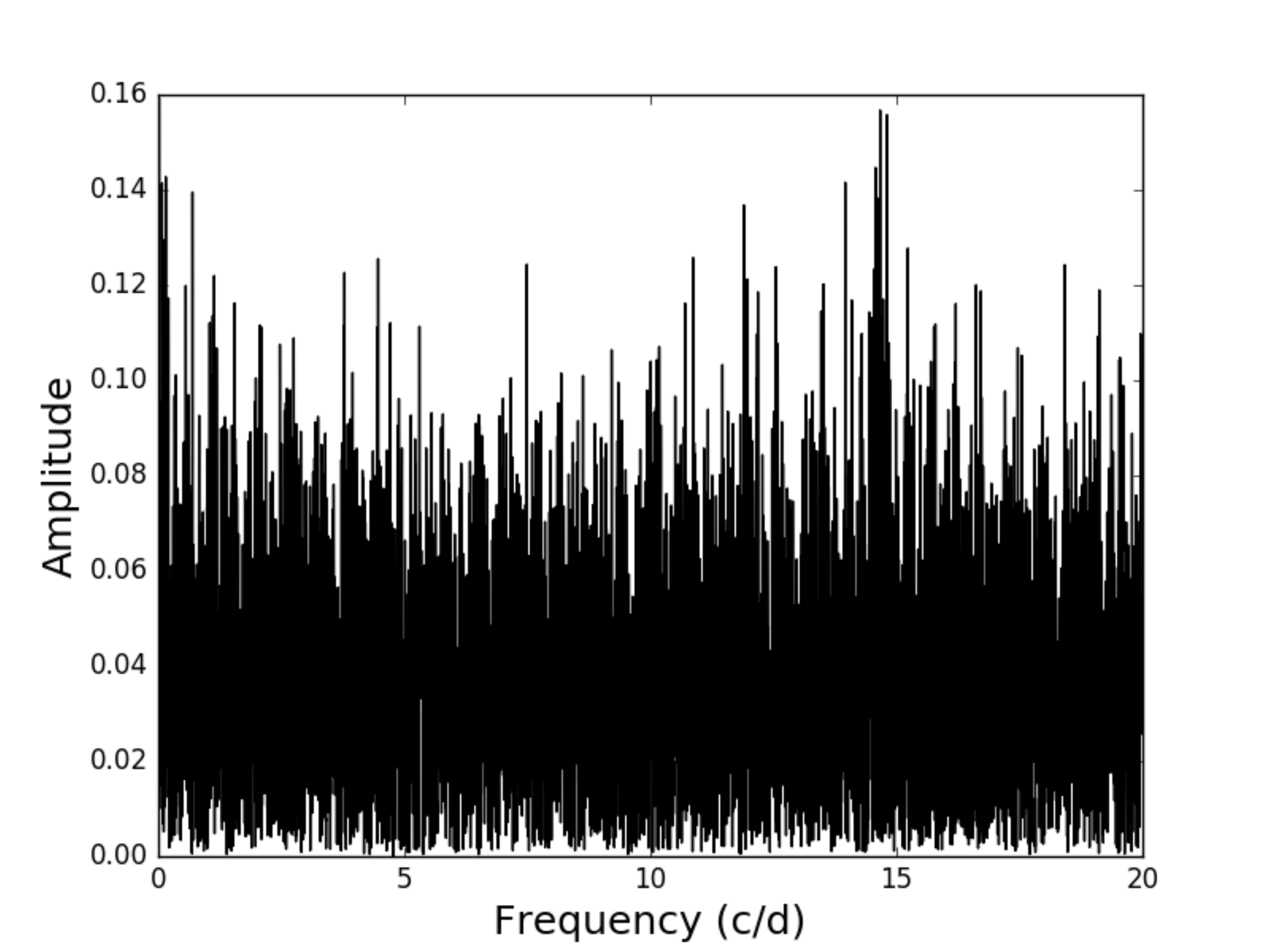}
    \caption{Fourier amplitude spectra of BTr data (in mmag). {\em Left -}\ Orbit-averaged data. {\em Right -}\ Spectrum of residuals following three iterations of prewhitening.}
    \label{fig1}
  \end{minipage}
\end{figure}

Fourier analysis with prewhitening was performed on the BRITE photometry using the {\sc Period04} package \citep{2005CoAst.146...53L}. One significant period was detected at $0.2095781(3)$~d, along with its first two harmonics. The original and prewhitened BTr Fourier amplitude spectra are illustrated in Fig.~\ref{fig1}. As is evident in Fig.~\ref{fig1}, the BTr photometry contains no evidence of signal beyond the modulation due to the fundamental period of $0.2095781(3)$~d and its first two harmonics. As a consequence, we conclude that the BRITE data reveal no evidence for new independent pulsation frequencies.

A similar analysis of the RV measurements by \citet{2017MNRAS.471.2286S} revealed a somewhat different situtation. Those authors found that prewhitening was unable to completely remove the comb associated with the pulsation. They ascribed this phenomenon to a changing pulsation period, and they concluded that the phenomenon is consistent with a period that grew linearly with time during the 17~yr of RV observations, at a rate of $0.0096(5)$~s/yr.

During the remainder of this paper we focus our attention on the issue of the evolving pulsation period, first to test the veracity of the claim, secondly to provide an independent determination of its rate of change, and finally to propose a physical interpretation of the phenomenon.

\section{Period evolution}

To evaluate the rate of period change $\dot{P}$, data were phased using an ephemeris that accounts for a linear period evolution (i.e. constant $\dot{P}$):

\begin{equation}
N = \frac{2\Delta{t}}{2P_0 + \dot{P}\Delta{t}}, \label{ephem}
\end{equation}

\noindent where $N$ is the number of cycles elapsed during the time $\Delta{t}$ (in days) since the temporal zero-point $T_0$, $P_0$ (in d) is the value of the period at time $T_0$, and $\dot{P}$ is the rate of period change. 

We began by analyzing the RV measurements reported by \citet{2017MNRAS.471.2286S}, for which we adopted the values of $T_0$ and reference period $P_0$ associated with the BRITE photometry.  

We first verified that the RV data could not be satisfactorily phased using a single, constant period. This was confirmed for both the combined (i.e. CORALIE + ESPaDOnS) dataset, as well as the individual datasets. 

We then used Eq.(\ref{ephem}) to phase the complete RV dataset for values of $\dot{P}$ ranging from $-0.02$ to $0.02$ s/yr in steps of $0.0001$ s/yr. For each value of $\dot{P}$, the phased RVs were fit using a second-degree harmonic function. By calculating the $\chi^2$ statistic for each  fit, we identified the value of $\dot{P}$ for which the dispersion of the measurements was minimized. We obtained a value of +0.0091 s/yr. The uncertainty was determined through a case resampling bootstrapping applied over 2500 iterations on the residuals to a harmonic fit to the data adopting the derived value of $\dot{P}$. The bootstrapping routine yielded an approximately normal distribution of values for $\dot{P}$ of which the standard deviation of $0.001$ was adopted as the $1\sigma$ uncertainty. 

\begin{figure}
  \centering
  \begin{minipage}{\textwidth}
    \includegraphics[width=0.45\textwidth]{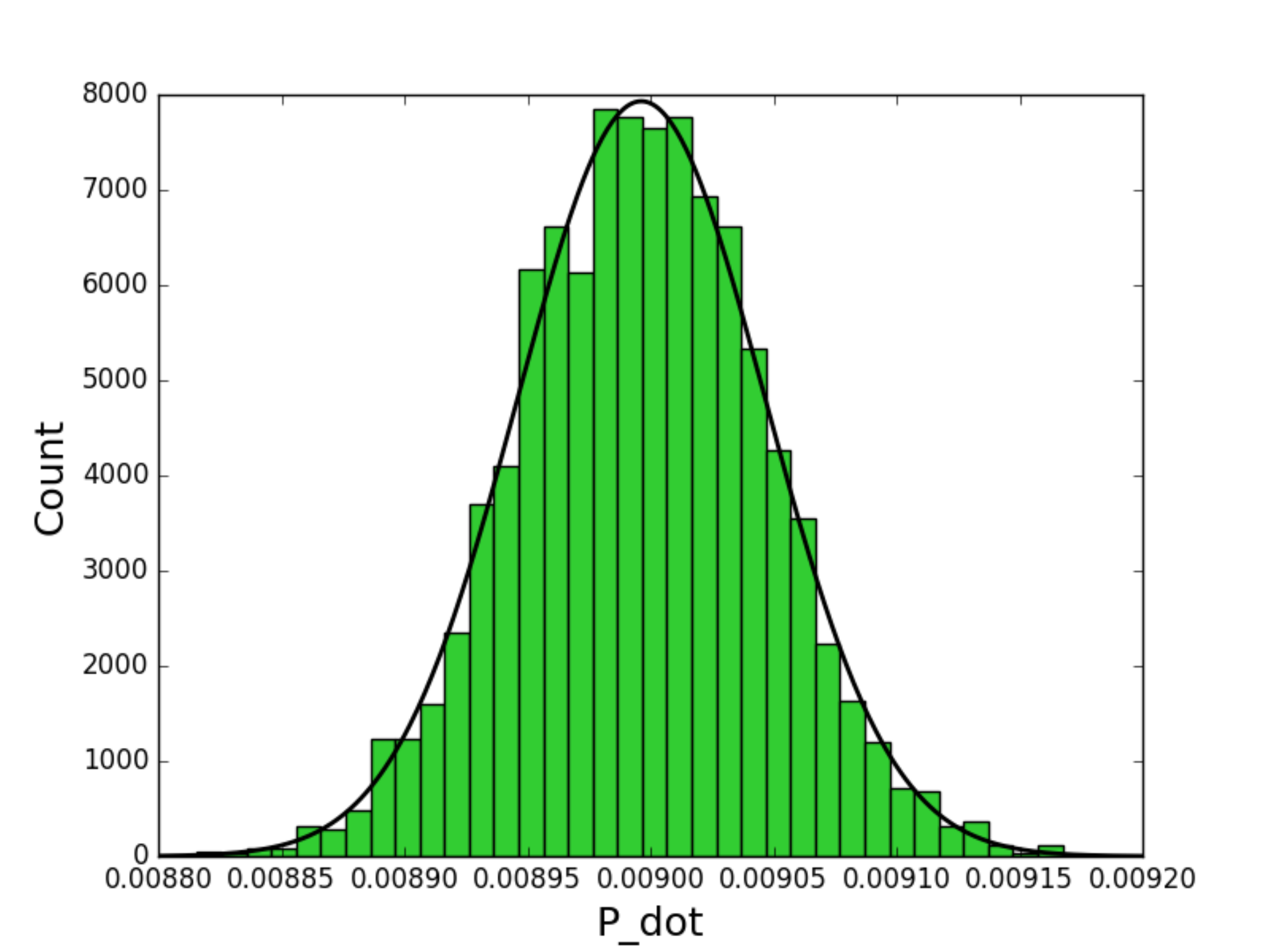}\includegraphics[width=0.45\textwidth]{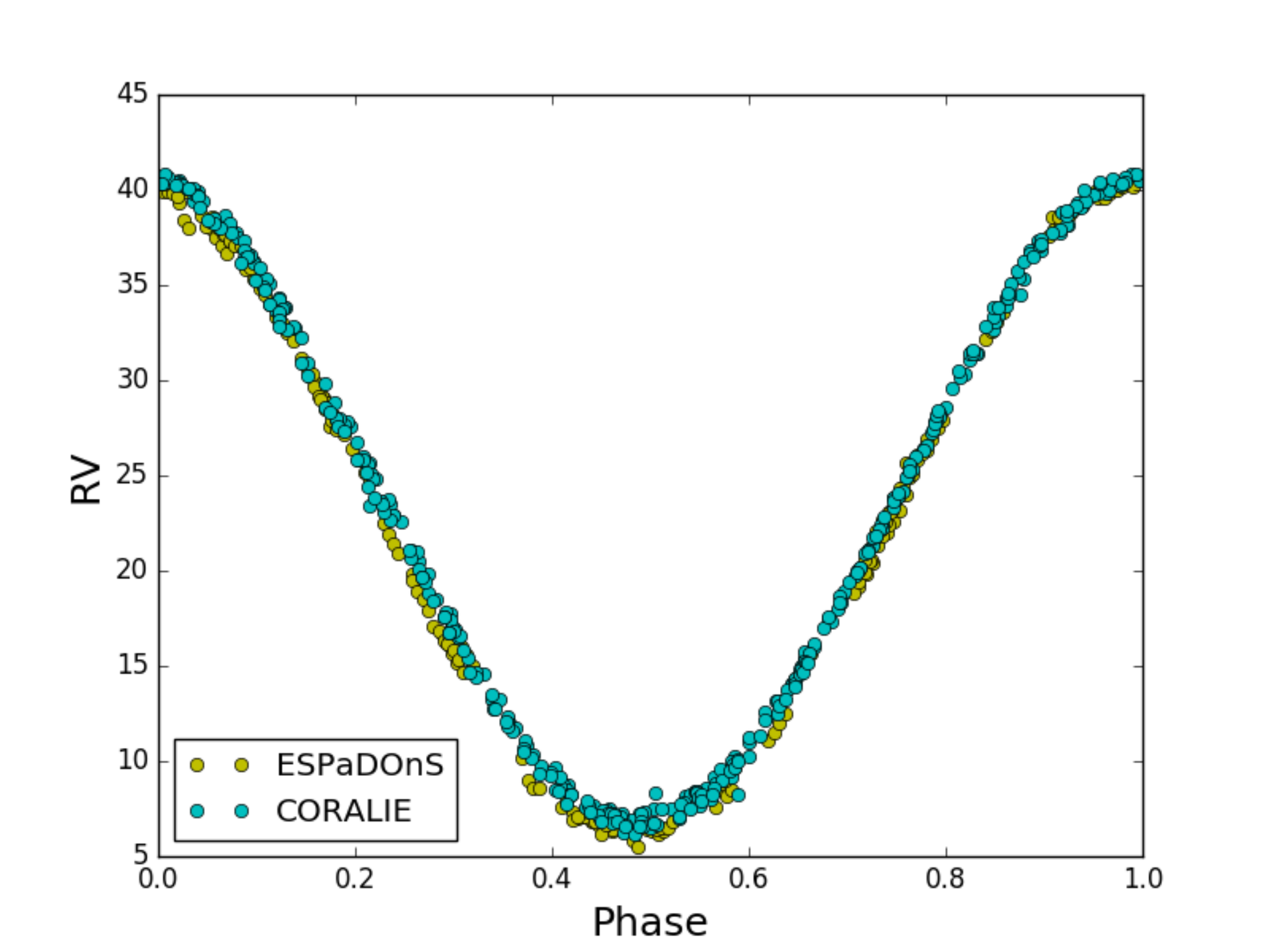}
    \caption{{\em Left -}\ Distribution of derived values of $\dot{P}$ inferred using the case resampling bootstrapping approach. {\em Right -}\ RV measurements of $\xi^1$~CMa phased according to Eq.~(\ref{ephem}), adopting the derived value of $\dot{P}$.}
    \label{pdots}
  \end{minipage}
\end{figure}

As a consequence of our analysis, we reported a value of $\dot{P}=0.009(1)$ s/yr, in agreement with the result of \citet{2017MNRAS.471.2286S}. The distribution of $\dot{P}$ values inferred as a result of our bootstrapping procedure, as well as examples of the RV curve phased using the derived value of $\dot{P}$, are shown in Fig.~\ref{pdots}.

As a final check, we performed the same analysis on the BRITE photometry. However, the much shorter duration of this timeseries resulted in a determination of $\dot{P}$ consistent with both zero and the value derived from the RVs. Combining the BRITE photometry with archival data will permit an independent test of our model.

\section{Interpretation}

The apparent evolution of the period of $\xi^1$~CMa may have several origins. Given that binarity is effectively ruled out based on the detailed analysis of \citet{2017MNRAS.471.2286S}, our focus in this paper is to interpret the growing period as a consequence of stellar evolution (e.g. \citealt{2015A&A...584A..58N}). As reported by those authors, the fractional rate of change of the pulsation period of a radially pulsating star due to evolving mass $M$ and radius $R$ on evolutionary timescales can be computed according to:

\begin{equation}
\frac{\dot{P}}{P} = -\frac{1}{2}\frac{\dot{M}}{M}+\frac{3}{2}\frac{\dot{R}}{R}. \label{pdot_evol}
\end{equation}

We have exploited the evolutionary model calculations of \citet{2012A&A...537A.146E} to predict the variation of $\dot{P}$ according to Eq.(\ref{pdot_evol}). Given that $\xi^1$~CMa is a (relatively) cool upper-main sequence star, its mass-loss rate is expected to be low; hence we have assumed $\dot{M}=0$ in Eq.(\ref{pdot_evol}).

\begin{figure}
  \centering
  \begin{minipage}{\textwidth}
    \includegraphics[width=0.45\textwidth]{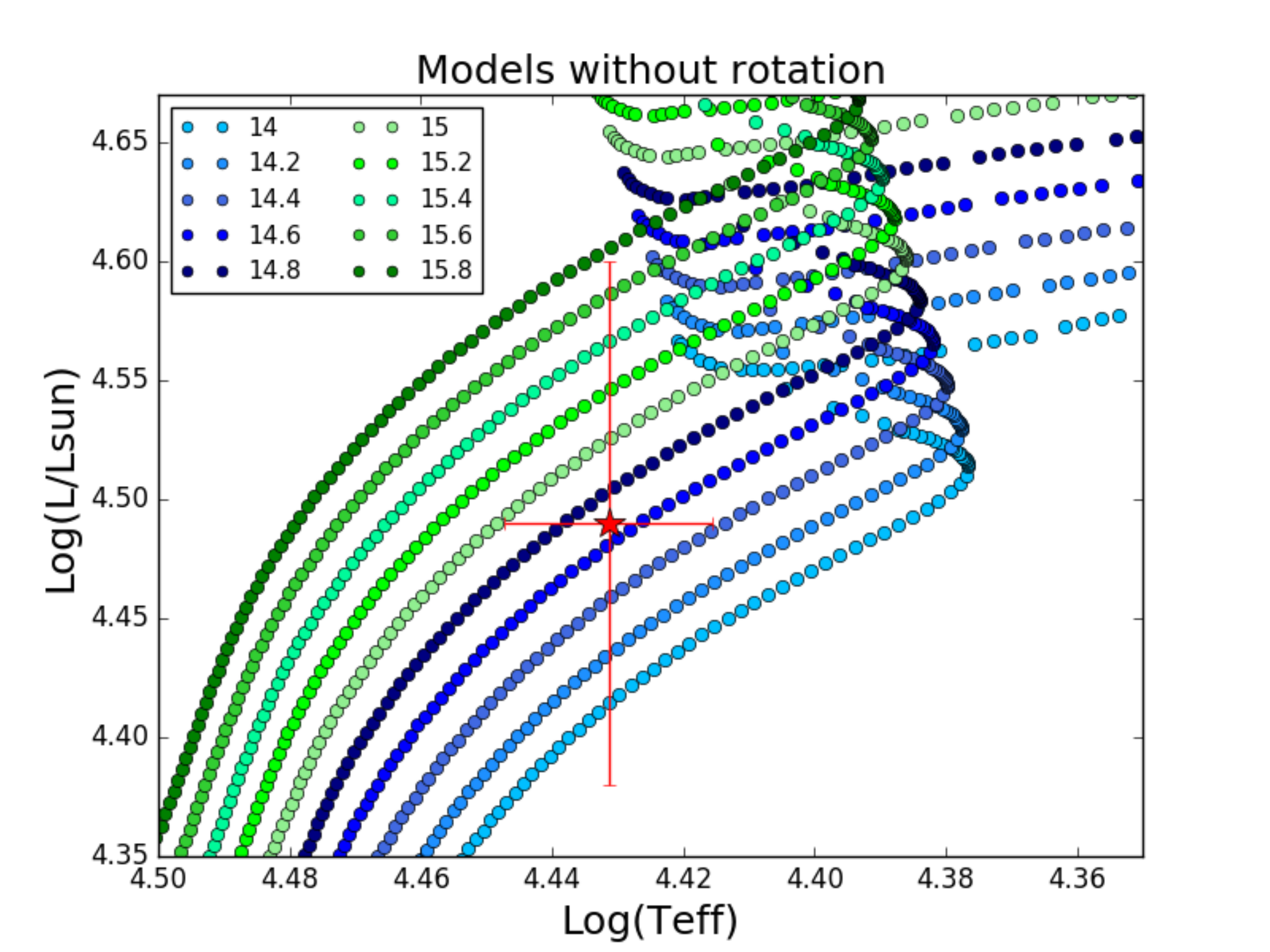}\includegraphics[width=0.45\textwidth]{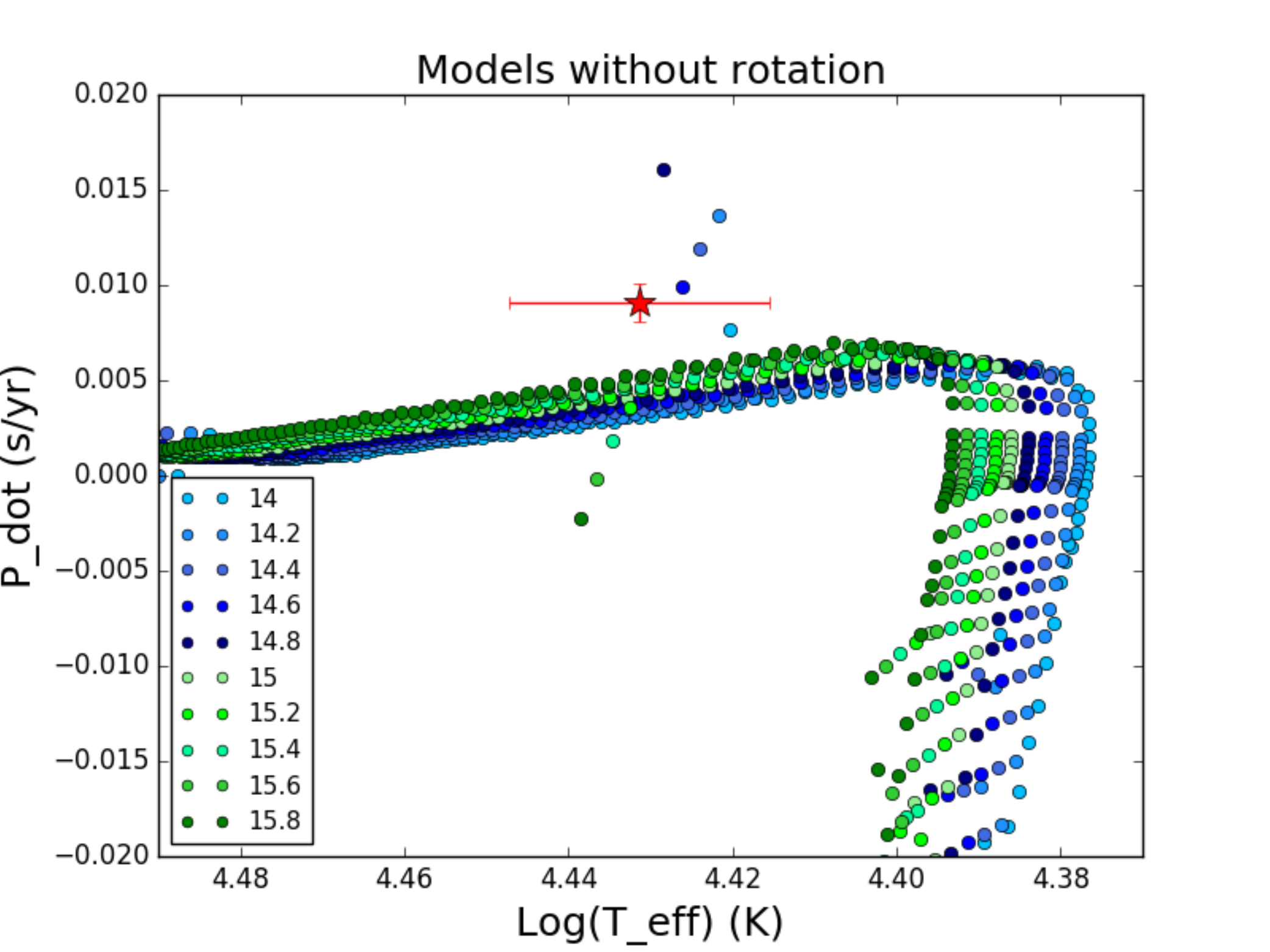}
        \includegraphics[width=0.45\textwidth]{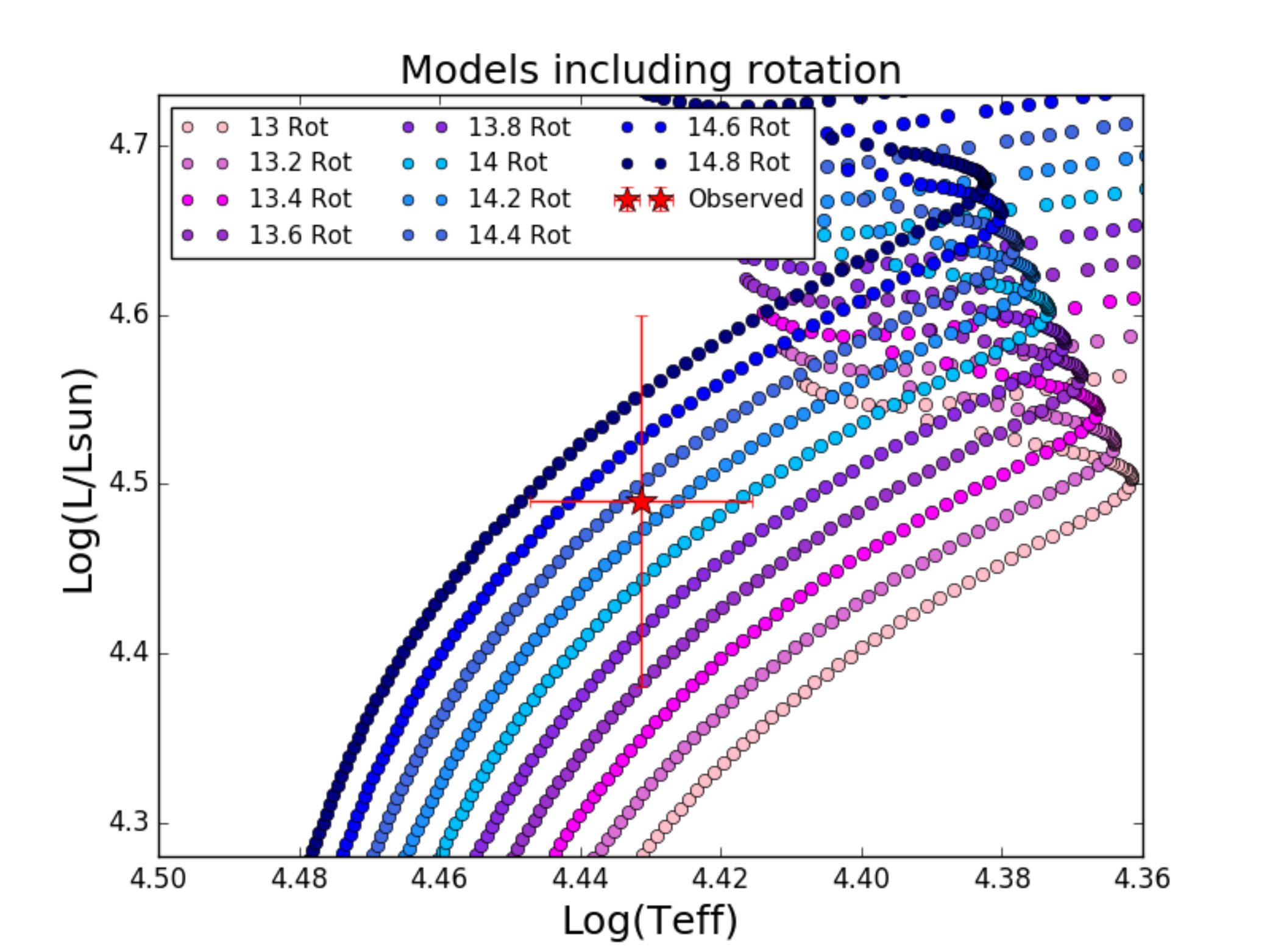}\includegraphics[width=0.45\textwidth]{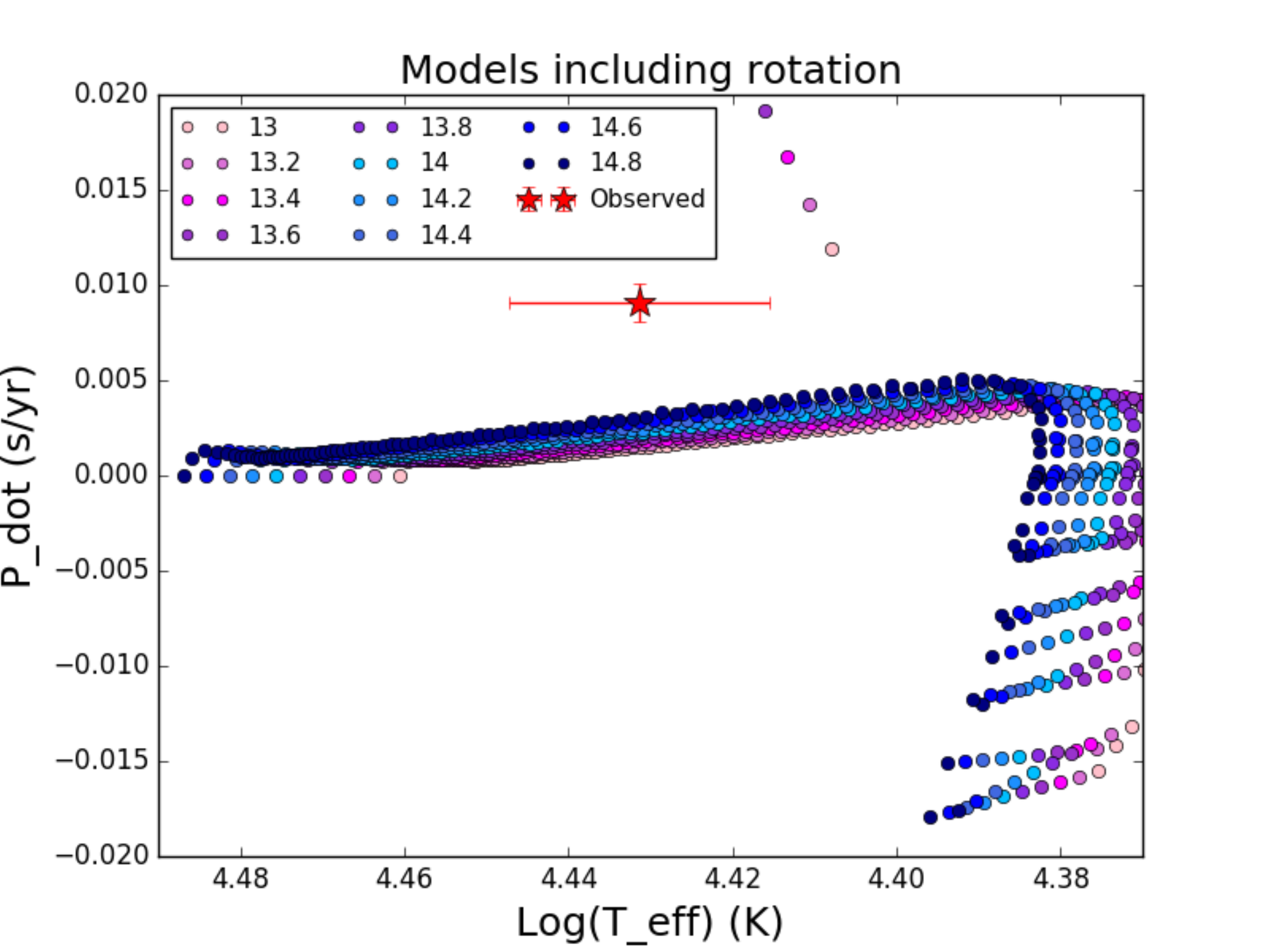}
    \caption{{\em Left -}\ Theoretical HR diagram showing evolutionary models ignoring (top) and including (bottom) the effects of rotation, calculated by \citet{2012A&A...537A.146E}. {\em Right -}\ $\dot{P}-T_{\rm eff}$ plane showing the same evolutionary models as at left, calculated according to Eq.(\ref{pdot_evol}) using the evolutionary models. The positions of $\xi^1$~CMa according to the physical parameters inferred by \citet{2017MNRAS.471.2286S} and the rate of period change determined here are shown in red.}
    \label{HRD}
  \end{minipage}
\end{figure}

The physical parameters of $\xi^1$~CMa determined by \citet{2017MNRAS.471.2286S} place the star on the second half of the main sequence. In Fig.~\ref{HRD} we show the star's position on the Hertzsprung-Russell (HR) diagram and on the $\dot{P}$ vs. $T_{\rm eff}$ diagram, for models both including the effects of rotation ($v_{\rm rot}=200$~km/s) and ignoring those effects, as computed by \citet{2012A&A...537A.146E}.  We find that only non-rotating models yield rates of period change formally consistent with the star's position on the HR diagram. However, the observed $\dot{P}$ is really only consistent with the predicted rate of change at precisely the terminal-age main sequence (TAMS). Given the specificity of the HR diagram location of the star required to satisfy this condition, further investigation of model sensitivities and systematics is required before any robust conclusions can be drawn.

\section{Conclusions}

\begin{itemize}
\item BRITE observations confirm $\xi^1$~CMa is monoperiodic pulsator exhibiting contributions to the light curves from the fundamental and first two harmonics. No additional independent frequencies are detected.
\item BRITE-Toronto measures the pulsation period of $\xi^1$~CMa to be $0.2095781(3)$~d.
\item Seventeen years of RV measurements reveal linear period change with $\dot{P}=0.009(1)$~s/yr, consistent with the results of \citet{2017MNRAS.471.2286S}.
\item This result is marginally consistent with the period evolution predicted by evolutionary models ignoring rotation, for a star located at the MS turnoff. The sensitivity of this result to the details of the evolutionary models will be a subject of future research.
\item  Combining the BRITE photometry with archival data will permit an independent test of the period evolution. 
\end{itemize}

\acknowledgments{Based on data collected by the BRITE Constellation satellite mission, designed, built, launched, operated and supported by the Austrian Research Promotion Agency (FFG), the University of Vienna, the Technical University of Graz, the Canadian Space Agency (CSA), the University of Toronto Institute for Aerospace Studies (UTIAS), the Foundation for Polish Science \& Technology (FNiTP MNiSW), and National Science Centre (NCN). AP acknowledges support from the NCN grant number 2016/21/B/ST9/01126. GH acknowledges support by the Polish NCN grants 2011/01/M/ST9/05914 and 2015/18/A/ST9/00578.}

\bibliographystyle{ptapap}
\bibliography{ptapapdoc}

\end{document}